\documentclass[fleqn,twoside]{article}
\usepackage{espcrc2}
\usepackage{graphicx}

\newcommand{\AmS}{{\protect\the\textfont2
  A\kern-.1667em\lower.5ex\hbox{M}\kern-.125emS}}

\title{Giant nonlinear response of superconducting single crystal niobium
    in a sweeping magnetic field.}

\author{M.I. Tsindlekht\address[HUJ]{The Racah Institute of Physics, The Hebrew University of Jerusalem,\\
    Jerusalem, 91904, Israel}%
    \thanks{The work at the Hebrew University was supported by 1998 Binational US-Israel Science Foundation
    grant and by the Klatchky foundation.}
    \thanks{e-mail address: mtsindl@vms.huji.ac.il},
        I. Felner\addressmark,
        G. Jung\address{Department of Physics, Ben Gurion University of the Negev, \\P.O.Box 653,
    84105 Beer-Sheva, Israel }
        \thanks{also with Instytut Fizyki PAN, Warszawa, Poland}}

\begin{document}

\begin{abstract}
Giant enhancement of the nonlinear response of a single crystal
of Nb placed in a sweeping magnetic field has been experimentally
observed. The rectified signal from Nb ($Tc=9.15$ K) has been
measured by means of an inductive method as a function of
temperature, dc field, dc field sweep rate, and the amplitude of
ac field. The Nb sample was excited by an amplitude modulated ac
field. Under a stationary regime, the rectified signal appears
only for magnetic fields ($H_0$) in the range $H_{c2}<H_0<H_{c3}$
. However, when the dc field was swept slowly, the rectified
signal appears at $H_0>H_{c1}$. This experiment shows that the
amplitude of the rectified signal is two orders of magnitude
larger than the amplitude of the signal seen under stationary
field conditions. Moreover, the amplitude of the rectified signal
is a power function of the sweep rate, with the power exponent
close to 1. \vspace{1pc}
\end{abstract}

\maketitle

\section{Introduction}

Non-stationary phenomena in type-II superconductors have attracted
attention since the middle sixties. Varying the dc magnetic field
results in an increase of the ac losses in the mixed state
\cite{MAX}. Moreover, while sweeping a dc magnetic field between
$H_{c1}$ and $H_{c2}$, a second harmonic signal appears in the
response to the ac field excitation \cite{CAMP}. H. J. Fink
\cite{FINK} showed that the second harmonic signal reaches
maximum at a sweep rate $dH_{{0}}/dt\approx \omega h_{{0}}$,
where $\omega $ and $h_{0}$ are the frequency and amplitude of ac
field respectively, and disappears when  $dH_{0}/dt<<\omega
h_{0}$.

Using high quality single crystal niobium, we have shown recently
that nonlinear response to small external ac excitation field
under stationary conditions, appears only in the surface
superconducting state \cite{TSIND}. In this paper we show that in
non-stationary conditions even non-linear effects appear also for
$H_{c1}<H_0<H_{c2}$. Moreover, the magnitude of this nonlinear
non-stationary response is orders of magnitude higher than the
response observed under stationary conditions \cite{TSIND} and,
therefore, it will be referred to as "a giant response".

\section{Experimental}
A rectangular sample, $10\times 3\times 1$ mm$^{3}$, was cut out
high purity single crystalline Nb ($T_c=9.15$ K and resistance
ratio $R_{300K}/R_{10K}\approx 300$). The results of magnetization
measurements of our sample were published elsewhere \cite{TSIND}.

The sample was exposed to dc $H_{0}$ and ac $h(t)$ magnetic
fields applied parallel to the [100] crystalline direction which
coincides with the longest dimension of the rectangular sample.
The amplitude modulated ac field $h(t)=h_{{0}}(1+\alpha cos\Omega
t)cos\omega t$, where: $0<h_{{0}}<0.6$ Oe, $\alpha \approx 0.9 $,
$\omega /2\pi =3.2$ MHz, and $\Omega /2\pi =733$ Hz, was obtained
by feeding the small primary copper coil from a high frequency
generator operating in a constant current regime. The detailed
description of the experimental setup was given elsewhere
\cite{TSIND}. Sweeping the dc magnetic field was achieved by a
normal metal solenoid of a commercial SQUID magnetometer system.
The maximum available amplitude of the field sweep was $120$ Oe
and sweep rate $dH_{{0}}/dt$ up to 24 Oe/s. After reaching the
maximum field at a given sweep rate, the field was decreased at
the same sweep rate. The nonlinear response has been measured
inductively by means of a secondary pick-up copper coil wound on
the primary coil.

Nonlinearity of a superconducting specimen immersed in an
amplitude modulated ac field results in oscillations of the
magnetic moment at the frequencies of the  harmonics of the
fundamental frequency $\omega$, and at the frequencies  $\omega
\pm \Omega $, as well as at the modulation frequency $\Omega $,
and its harmonics. The pick-up coil transforms the oscillations
of the magnetic moment into ac voltage signals. The signal was
processed by a lock-in amplifier as a function of the experimental
parameters such as: temperature, $H_0$, dc, $dH_{0}/dt$, and ac
field amplitude $h_{0}$.

The nonlinear response to a dc magnetic field sweep, i.e. the
amplitude of the rectified signal $ A_{\Omega }$ at the frequency
$\Omega $, has been measured as follows. The sample was first
cooled down to the required temperature at zero magnetic field.
Next, $H_0$ was raised to $H_{st}$, the actual sweep of $H_0$ was
ramped-up, and eventually $ A_{\Omega }$ was measured as a
function of time. In the following we shall discuss only the
phenomena associated with the increasing magnetic field sweep.
The hysteretic behavior observed when the field was swept down
will be discussed elsewhere.

\section{Results and discussion}

 Figure 1 shows the field dependence of the rectified signal
$ A_{\Omega}$ measured at T=8.5 K for various $H_{st}$ values.
Both sweep rate and ac amplitude $h_{0}$ are constant. It is
readily observed that for $H_0<H_{c1}$ the response is negligible.
However, for $H_0>H_{c1}$ a slight increase in the signal
appears. For $H_{0}$ close to $H_{c2}$ the magnitude of the
response grows significantly, and exceeds the stationary response
measured in the same setup with the same sample by two orders of
magnitude\cite{TSIND}. This result contradicts previous
experimental observation \cite{MAX,CAMP} and theoretical
predictions\cite {FINK}. This is due to the fact that in our
experiments $dH_{0}/dt<<\omega h_{0}$ and therefore, in
accordance with Fink \cite{FINK}, $A_{\Omega }$ should be zero.
\begin{figure}[hb]
\includegraphics[width=15pc]{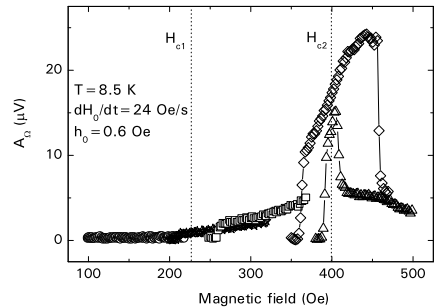}
\vskip -0.8truecm \caption{Rectified signal $ A_{\Omega }$ vs. sweeping
magnetic field at 8.5 K. Different symbols correspond to different starting
fields $H_{st}$. Dotted lines indicate positions of $H_{c1}=225$ Oe and
$H_{c2}=400$ Oe.} \label{fig1}
\end{figure}

Further increase of $H_{st}$ towards $H_{c2}$ leads to a new
phenomenon. Rapidly increasing $A_{\Omega}$ with a the maximum at
$H_0>H_{c2}$ sharply drops at $H_0=H_{k}$. For $H_0>H_k$
$A_{\Omega}$ remains almost constant for any $H_{st}$ and then
the signal gradually decreases to zero with $H_0$ approaching
$H_{c3}\approx 600$ Oe. At the same time all hysteretic phenomena
disappear when $H_{st}\geq H_{c2}$.

The value of $H_k$ at which $A_{\Omega}$ abruptly drops turned
out to be strongly dependent on the experimental parameters such
as $H_{st}$, $h_0$ and $dH_0/dt$. We have found that with
decreasing the sweep rate and/or with increasing the excitation
amplitude and starting field, the characteristic field $H_k$
asymptotically approaches $H_{c2}$. For example, Fig.~2 shows the
dependence of $H_k$ on sweep rate $dH_0/dt$. The solid line is a
best fit to $H_k\propto (dH_0/dt)^{2}$.
\begin{figure}[h]
\includegraphics[width=15pc]{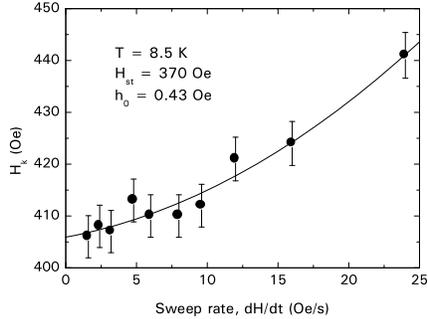}
\vskip -0.8truecm \caption{Characteristic field $H_k$ at which the
rectified signal abruptly drops as a function of the sweep rate of the
magnetic field.} \label{fig2}
\end{figure}
We would like to emphasize that Fig.~2 presents some type of
dynamical phase diagram. Namely it is clear that the solid line
is a dynamic boundary between surface and bulk superconductivity.

The experiments revealed  that the amplitude of the rectified
signal increases with $dH_{0}/dt$. This dependence, determined
for three different values of the $H_0$, is presented in Fig.~3.
The power law fit to the data shows that $A_{\Omega }$ $\propto
\left( dH_{0}/dt\right)^{p}$ with the exponent $p=0.95 \pm 0.02$.
The exponent $p$ does not depend on temperature, $h_{0}$ and
$H_{st}$. However as $H_{0}$ approaches the field at which
$A_{\Omega}$ goes through a maximum, the exponent $p$ goes down,
as illustrated in Fig.~3 for $H_0=415$ Oe, very close to the
response maximum, Fig.~1.
\begin{figure}[ht]
\includegraphics[width=15pc]{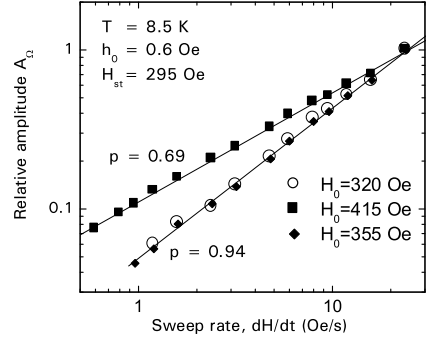}
\vskip -0.8truecm \caption{Amplitude of the rectified signal
normalized to a maximum amplitude at a given $H_{0}$ as function
of magnetic field sweep rate determined at three different values
of the magnetic field $H_0$, see the legend.} \label{fig3}
\end{figure}

The discrepancy between the  observed phenomena and previously
reported results can be understood considering a sand pile
analogue of a  critical vortex state \cite{PG}. Slow increase of
the dc field leads to an avalanche like vortex array
redistribution. All previous experiments were performed at low
frequencies at which any modifications of the nonlinear response
due to vortex avalanches cannot be observed since at low
frequencies $dB/dt\gg \omega h_0$. At high frequencies vortex
avalanches become effective. Namely at high frequencies $dB/dt
\approx \omega h_0$ is possible. Moreover, vortex avalanches
occur in the entire volume of the sample and not only in the
surface layer. The increase of the active volume may result in a
significant enhancement of the nonlinear response. The proposed
model may qualitatively explain the appearance near $H_{c2}$ of a
sharp drop of the rectified signal at $H_k$.  It is clear, that
only a quantative theory will be able explain features observed
experimentally in the present work. However, to the best our
knowledge this theory does not exist as yet.

\end{document}